\documentclass[preprint,12pt,tighten]{aastex}

\usepackage{graphicx}
\usepackage{amsmath}

\begin{document}

\voffset -0.5in
\textheight 9.truein
\let\oldthebibliography=\thebibliography
  \let\endoldthebibliography=\endthebibliography
  \renewenvironment{thebibliography}[1]{%
    \begin{oldthebibliography}{#1}%
      \setlength{\parskip}{0ex}%
      \setlength{\itemsep}{0ex}%
  }%
  {%
    \end{oldthebibliography}%
  }


\title{Detection of Dark Matter Decay in the X-ray}

\title{\small Submitted to the Astro2010 Decadal \\ 
Cosmology and Fundamental Physics Science Frontier Panel}
 
\author{Kevork N. Abazajian} \affil{Maryland Center for Fundamental
  Physics, Department of Physics, University of Maryland, College
  Park, MD 20742, USA}

\begin{abstract}
  There exists considerable interest in dark matter candidates that
  can reduce cosmological structure at sub-galactic scales through a
  suppression of the power spectrum of primordial perturbations as
  well as have a primordial velocity distribution to produce cores in
  the smallest-scale dwarf halos.  Light keV-mass-scale sterile
  neutrinos can be such a ``warm'' dark matter candidate and are still
  viable candidates in broad regions of their parameter space.  We
  review the status of this candidate and the current constraints from
  its radiative decay in the X-ray and from structure formation.  We
  also provide a forecast for the sensitivity of the {\it
    International X-ray Observatory} to this decay, which we show may
  have the ability to detect or exclude sterile neutrinos within its
  full parameter space of interest as a dark matter candidate.
\end{abstract}

\maketitle

\pagenumbering{arabic}

\section{Introduction}

The nature of dark matter remains one of the most significant unsolved
problems in cosmology and particle astrophysics.  The first indication
of the presence of what was dubbed ``dark matter'' was by Fritz
Zwicky, who in 1933 suggested a dark matter component to clusters of
galaxies to explain the high mass-to-light ratios required by
measurements of high velocity dispersions of component
galaxies~\citep{zwicky1933}.  It was then found that the rotation
curves of galaxies also required a dark matter component.  The
formation of large-scale structure measured by galaxy surveys and the
cosmic microwave background (CMB) also has required the presence of
dark matter.  In fact, the abundance of dark matter has been precisely
determined by observations of anisotropies in the CMB
\citep{Komatsu:2008hk}, and from the growth of cosmological structure
in the clustering of galaxies~\citep{Tegmark:2003ud}.  The fundamental
nature of the dark matter, however, remains unknown.

One natural candidate is a fermion that has no standard model
interactions other than a coupling to the standard neutrinos through
their mass generation mechanism~\citep{Dodelson:1993je,Shi:1998km}.
Due to their lack of interactions and association with the neutrino
sector, such fermions are referred to as sterile neutrinos.  One
prominent search for sterile neutrinos was the MiniBooNE experiment,
which searched for oscillations consistent with a form of sterile
neutrino indicated an oscillation interpretation of the Liquid
Scintillator Neutrino Detector (LSND) results, at a mass scale of
$m_s \sim 1\rm\ eV$ and mixing of $sin^2 2\theta\sim 10^{-2}$.  MiniBooNE found
evidence against such a sterile neutrino interpretation for
LSND~\cite{AguilarArevalo:2007it}, which is very far from the
mass and mixing scale required for dark matter sterile neutrinos,
$m_s \sim 1\rm\ keV$ and mixing $sin^2 2\theta \lesssim 10^{-7}$.

Observations are consistent with sterile neutrinos as the dark matter
for a limited mass range for the standard production mechanism.  The
most stringent constraints come from X-ray observations and
observations small scale cosmological
structure~\citep{Abazajian:2006yn}.  In this allowed range of masses,
the sterile neutrino has a non-negligible thermal velocity component,
and is therefore a warm dark matter (WDM) candidate, but for higher
particle masses, can behave as a cold dark matter (CDM) particle
candidate.  Much work has been done on oscillation production of this
dark matter candidate, which ties its production mechanism and
primordial density with its decay rate and
detectability~\citep{Abazajian:2001nj,Abazajian:2002yz,Asaka:2006rw}.
Other production mechanisms have the sterile neutrino production
method separate from the physics of their decay, and therefore could
lie in any part of the detectable parameter
space~\citep{Kusenko:2006rh,Shaposhnikov:2006xi}.  As shown below,
{\it IXO} can detect this dark matter candidate in the entire favored
region of parameter space or rule it out, using techniques proven by
current and past observations.

The prevalent ansatz of an absolute CDM component in galaxy formation
is not strictly valid even for one of the most cited CDM candidate, a
supersymmetric neutralino.  The damping scale at which thermal
velocities of the dark matter cut off the growth of gravitationally
bound structures is completely unknown below the galactic scale.  One
principal challenge to the CDM paradigm is the order of magnitude
over-prediction of the observed satellites in galaxy-sized halos such
as the Milky Way \citep{Klypin:1999uc,Moore:1999wf}.  Warm dark matter
suppresses dwarf galaxy formation, which may occur through
fragmentation of larger structures.  Semi-analytic galaxy formation
modeling has found that the number of dwarf galaxies formed in
satellite halos may be suppressed due to the reionization, stellar
feedback within halos, and/or tidal stripping of satellites
\citep{Bullock:2000wn}.   Whether a minor or major suppression of small
mass halos is beneficial or detrimental to the suppression of dwarf
galaxy formation remains unsolved.

Four more problems in the CDM paradigm may benefit from the reduction
of power on small scales from WDM. First is the reduction of the
prevalence of halos in low-density voids in N-body simulations of CDM
structure formation, consistent the apparent dearth of massive
galaxies within voids in local galaxy
surveys~\citep{Peebles:2001nv}. The second is the relatively low
concentrations of galaxies observed in rotation curves compared to
what is predicted from the $\Lambda$CDM power
spectrum~\citep{Dalcanton:2000hn}, which can be relieved by a
reduction of the initial power spectrum of density fluctuations at
small scales~\citep{Zentner:2002xt}. The third is the
``angular-momentum'' problem of CDM halos, where gas cools at very
early times into small mass halos and leads to massive low-angular
momentum gas cores in galaxies, which can be alleviated by the
hindrance of gas collapse and angular momentum loss through the delay
of small halo formation in a WDM scenario~\citep{Dolgov:2001nq}.  The
fourth problem is the formation of disk-dominated or pure-disk
galaxies in CDM models, which is impeded by bulge formation due to the
high merger accretion rate history in CDM models, but may be
alleviated with WDM~\citep{Governato:2002cv}.

Of particular interest recently for models of WDM are the possible
indications of the presence of cores in local group dwarf galaxies,
inferred from the positions of central stellar globular
clusters~\citep{Goerdt:2006rw} and radial stellar velocity dispersions
\citep{Wilkinson:2006qq}.  For example, for the Fornax dwarf
spheroidal galaxy to be consistent with a packed phase space density
of the WDM, as a sterile neutrino, the sterile neutrino particle mass
must be between 0.5 and 1.2 keV
\citep{Strigari:2006ue,Abazajian:2006yn}.  Conversely, this places a
robust limit---the Tremaine-Gunn bound---on the mass and phase space
of the dark matter particle from observed dynamics in galaxy
centers~\citep{Tremaine:1979we}.

\begin{figure}
  \centerline{\includegraphics[width=4.8truein]{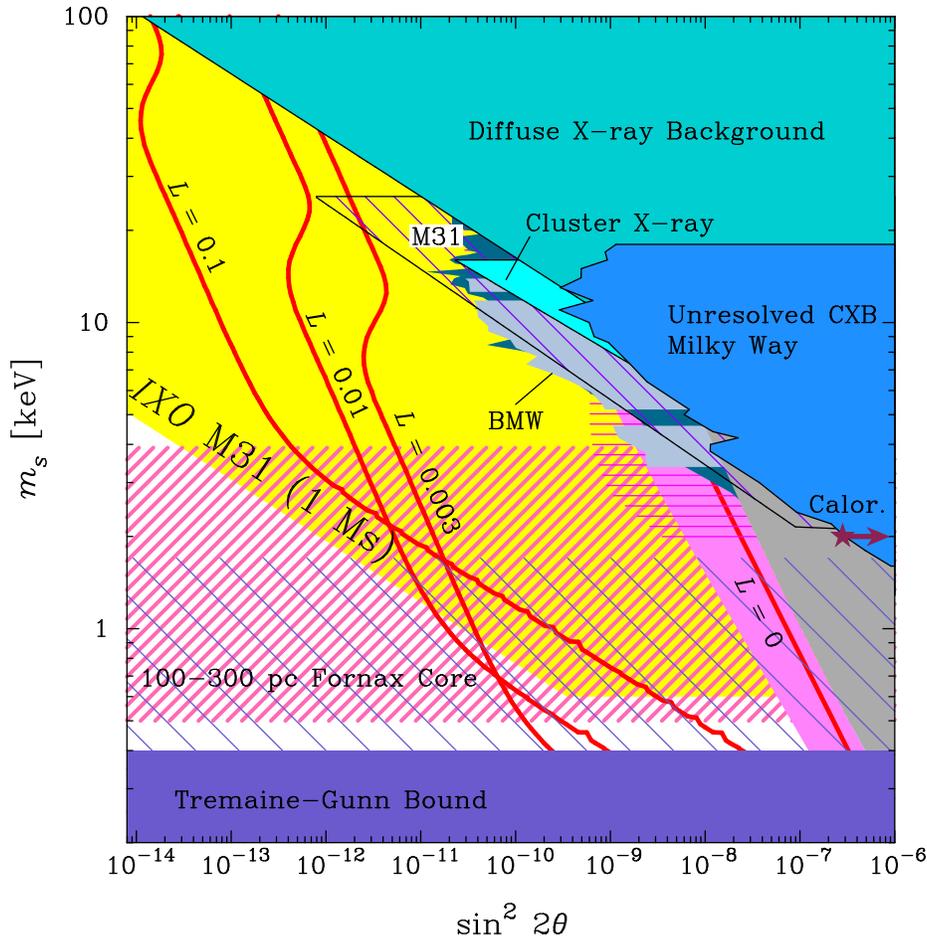}}
  \figcaption{\footnotesize Full parameter space constraints for the
    sterile neutrino production models, assuming sterile neutrinos
    constitute the dark matter.  Favored regions are in red/magenta
    colors, disfavored and excluded regions are in blue/turquoise
    colors.  The strongly favored region of a horizontal band of the
    mass scale consistent with producing a 100 - 300 pc core in the
    Fornax dwarf galaxy~\citep{Strigari:2006ue}. The favored
    parameters consistent with pulsar kick generation are in
    horizontal hatching~\citep{Kusenko:2008gh}. Contours labeled with
    lepton number $L=0$, $L=0.003$, $L=0.01$, $L=0.1$ are dark matter
    production predictions for constant comoving density of
    $\Omega_s=0.24$ for $L=0$, and $\Omega_s=0.3$ for non-zero lepton
    number $L$ universes~\citep{Abazajian:2002yz}.  The minimal
    standard production prediction is that of $L=0$.  The grey region
    to the right of $L=0$ over-produces the dark matter.  Constraints
    from X-ray observations include the diffuse X-ray background
    (turquoise)~\citep{Boyarsky:2005us}, from XMM-Newton observations
    of the Coma and Virgo clusters (light
    blue)~\citep{Boyarsky:2006zi}, the Milky Way (labeled
    BMW)~\citep{Boyarsky:2006ag}, and Ursa Minor using Suzaku (dark
    blue)~\citep{Loewenstein:2008yi}.  The diagonal wide-hatched
    region is the constraint from {\it XMM-Newton} observations of
    M31~\citep{Watson:2006qb}.  Also shown is the best current
    constraint from {\it Chandra}, from observations of contributions of
    dark matter X-ray decay in the cosmic X-ray background through the
    CDFN and CDFS (medium blue, ``Unresolved CXB Milky Way'').  Also
    shown is an estimate of the sensitivity of a 1 Ms observation of
    M31 with {\it IXO} (yellow).  This extends into the entire favored
    parameter space of interest where all production models can place
    the dark matter candidate.  The region at $m_s<0.4\rm\ keV$ is
    ruled out by a conservative application of the Tremaine-Gunn
    bound~\citep{Bode:2000gq}, and $m_s<1.7\rm\ keV$ is disfavored by
    conservative application of constraints from the Lyman-$\alpha$
    forest.
  \label{parameterspace}}
\end{figure}

There are three other interesting physical effects when sterile
neutrinos have parameters such that they are created as the dark
matter in the non-resonant production mechanism.  First, asymmetric
sterile neutrino emission from a supernova core can assist in
producing the observed large pulsar velocities above $1000\rm\ km\
s^{-1}$ \citep{Kusenko:1998bk,Kusenko:2008gh}.  The parameter space
overlaps that of the non-resonant production mechanism
(Fig.~\ref{parameterspace}). Second, the slow radiative decay of the sterile neutrino
dark matter in the standard production mechanism can augment the
ionization fraction of the primordial gas at high-redshift
(high-$z$)~\citep{Biermann:2006bu}.  This can lead to an enhancement
of molecular hydrogen formation and star formation.  This has
significant consequences on the formation of the first stars and
quasars, leading to earlier formation of massive gas systems, stars
and quasars than expected in WDM or even canonical CDM models.  This
remains an open question, but it does indicate that the reionization
epoch is not a elementary constraint on WDM
models~\citep{O'Shea:2006tp}.  Third, the sterile neutrino may enhance
the heating of the shock in core-collapse Type-II supernovae,
alleviating the problems of ~\citep{Hidaka:2007se}.

The potentially beneficial effects of the suppression of cosmological
small-scale structure in WDM models like sterile neutrinos can also
lead to observational conflicts if the suppression extends to
excessively high mass and length scales.  The suppression scale
monotonically decreases with increasing sterile neutrino particle
mass.  One of the best direct measures of clustering at small scales
is the clustering observed in intervening gas along the line-of-sight
to a quasar, known as the Lyman-$\alpha$
forest~\citep{Narayanan:2000tp}.  Statistically-consistent constraints
allowing freedom in all cosmological parameters and constraints from
the CMB, galaxy clustering, and a measurement of clustering in the
Lyman-$\alpha$ forest gives a lower limit for the sterile neutrino
dark matter particle mass as $m_s >1.7\rm\ keV
\ (95\%\ CL)$~\citep{Abazajian:2005xn}.  \citet{Seljak:2006qw}
describe a much more stringent constraint when directly using high-$z$
flux power spectra from the Sloan Digital Sky Survey (SDSS) and other
higher-resolution flux power spectra, $m_s >14\rm\ keV \ (95\%\ CL)$.
This limit only applies to the specific non-resonant oscillation
production model of \citet{Dodelson:1993je} and not to other
production methods such as non-zero lepton number $L$ cosmologies.
Moreover, the analysis and interpretation of the Lyman-$\alpha$ forest is
complicated by many aspects.  The theoretical interpretation of the
flux power spectrum is complicated by the thermal state of the gas as
a function of redshift, the presence of metal lines in the forest, and
the limitations of hydrodynamical simulations in the forest modeling
\citep[e.g.][]{Jena:2004fc}.

There remains considerable interest in the potential
benefits of WDM in galaxy and sub galaxy-scale structure formation,
and sterile neutrinos are the most natural candidate.  Therefore,
there is still considerable interest in searching for the dark matter
decay line in the X-ray.  

\section{Detectability by X-ray Observations}

Because the sterile neutrino dark matter candidate is mixed with one
or more active neutrinos, there exists are radiative decay mode of the
dark matter particle to the lighter neutrino and an X-ray photon.
\cite{Abazajian:2001vt} was the first to propose the use of X-ray
observations to detect this signature radiative decay.  Since the
mixing angle required for production of the sterile neutrino dark
matter is such that $\theta^2 \lesssim 10^{-6}$, the coupling is
inaccessible to any current methods of searches for neutral leptons
via so-called kink-searches in beta-decay experiments.  However, the
same mass-generation mechanism that couples the sterile neutrino to
the active neutrinos for their production causes a radiative decay
mode that can be detected by X-ray observatories.  For the Majorana
neutrino case, the decay rate is \citep{Pal:1981rm}
\begin{equation}
\Gamma_\gamma(m_s,\sin^2 2\theta) \approx 1.36 \times 10^{-30}\,{\rm
s^{-1}}\ \left(\frac{\sin^2 2\theta}{10^{-7}}\right)
\left(\frac{m_s}{1\,\rm keV}\right)^5,
\end{equation}
where $m_s$ is the mass eigenstate most closely associated with the
sterile neutrino, and $\theta$ is the mixing angle between the sterile
and active neutrino.  The decay of a nonrelativistic sterile neutrino
into two (nearly) massless particles produces a line at energy
$E_\gamma = m_s/2$.  The radiative decay time is orders of magnitude
greater than the age of the universe (which is necessary for a viable
dark matter candidate) but the number of particles in the field of
view of the {\it Chandra} or {\it IXO} observatories is
approximately $\sim 10^{70}$, which makes this decay even at very low
rates detectable by these observatories.

Since the initial proposal of the search of the radiative decay line
in clusters of galaxies, field galaxies, and the cosmic X-ray
background by \citet{Abazajian:2001vt}, many groups have conducted
follow-up searches for the signature decay in the cosmic X-ray
background \citep{Boyarsky:2005us}, clusters of galaxies
\citep{Watson:2006qb}, dwarf galaxies \citep{Boyarsky:2006fg}, the
Andromeda galaxy \citep{Watson:2006qb} and the Milky Way
\citep{Riemer-Sorensen:2006fh}.  No detections of candidate lines have
yet been found, but upper limits to the decay flux in lines have been
made, and have led to upper limits to the particle mass of the sterile
neutrino dark matter in the standard production mechanism.  The best
current constraint is that from an analysis of {\it XMM-Newton} X-ray
observations of the Andromeda galaxy by \citet{Watson:2006qb}, with
the constraint at the level of $m_s < 3.5\rm\ keV\ (95\%\ CL)$.
Constraints from Milky Way observations are comparable in
strength~\citep{Boyarsky:2006ag}, as well as constraints from {\it
  Suzaku} observations of Ursa Minor~\citep{Loewenstein:2008yi}.  Many
of the current X-ray constraints on the parameter space of sterile
neutrino dark matter parameters are shown in Fig.~1.

\section{Sensitivity of the {\it IXO} Telescope and Detectors}

{\it IXO} has very significant technical advantages to the detection
of a dark matter decay line over current observatories such as {\it
  Chandra}, {\it XMM-Newton}, or {\it Suzaku}.  The advantages come
from three factors: (1) the significantly larger collecting area
of{\it IXO}, (2) the enhanced energy resolution of the {\it IXO}
microcalorimeter detectors, and (3) lower background of the {\it IXO}
spectral and imaging system.


The primary advantage for this application of {\it IXO} over that of
current observatories is its collecting area, which increases the
sensitivity of the detection by a factor of approximately $75$ times
relative to {\it Chandra}.  The second advantage is the planned
microcalorimetric energy resolution of $2.5$ eV, which enhances the
sensitivity of the decay line over background relative to that aboard
{\it Chandra} by a factor of approximately $8.9$.  The third advantage
is the planned significant reduction of background radiation in the
spectral and imaging system, which enhances the sensitivity to a line
feature by a factor of approximately $12$ times relative to {\it
  Chandra}.\footnote{Scientific and technological design goal data is
  from {\tt http://ixo.gsfc.nasa.gov/} and the {\it IXO} Science Team
  Meeting January 28-29, 2009.}

In detail, it can be shown that the background count rate in a line of
$2.5$ eV in {\it IXO} is approximately $B \approx 1.9 \times 10^{-5}
t_5\rm\ counts$ in an integration time of $t_5 = t/(10^5\rm\ s)$.  The
count rate in a line is $C_L = 2.3 \times 10^{-2}
F_{-14}\rm\ counts\ s^{-1}$ for a line of flux $F_{-14} \equiv
F/(10^{-14}\rm\ ergs\ cm^{-2}\ s^{-1}$.  The count level sensitivity for
a $4\sigma$ line is then $C_L = 4\sqrt{B}/t$, or the detectable flux
is $F^{\rm det}_{-14} \approx 9.3\times 10^{-6}t_5^{-1/2}$.

The sensitivity detailed here is a several order of magnitude
improvement to line detection from currently operational X-ray
observatories.  The sensitivity in parameter space of {\it IXO} to a
sterile neutrino-like dark matter particle is shown in Fig.~1 relative
to current constraints.  As shown in the figure, the entire favored
region is within reach for {\it IXO}.  In summary, {\it IXO} would
detect this dark matter particle if it exists within any of the
favored parameter space (colored red/magenta), or exclude it entirely,
which presents a tremendous opportunity for particle astrophysics and
cosmology within our lifetime.

\bibliography{sndm}

\begin{thebibliography}{43}
\expandafter\ifx\csname natexlab\endcsname\relax\def\natexlab#1{#1}\fi

\bibitem[{Abazajian(2006)}]{Abazajian:2005xn}
Abazajian, K. 2006, Phys. Rev., D73, 063513, astro-ph/0512631

\bibitem[{Abazajian {et~al.}(2001{\natexlab{a}})Abazajian, Fuller, \&
  Patel}]{Abazajian:2001nj}
Abazajian, K., Fuller, G.~M., \& Patel, M. 2001{\natexlab{a}}, Phys. Rev., D64,
  023501, astro-ph/0101524

\bibitem[{Abazajian {et~al.}(2001{\natexlab{b}})Abazajian, Fuller, \&
  Tucker}]{Abazajian:2001vt}
Abazajian, K., Fuller, G.~M., \& Tucker, W.~H. 2001{\natexlab{b}}, Astrophys.
  J., 562, 593, astro-ph/0106002

\bibitem[{Abazajian \& Koushiappas(2006)}]{Abazajian:2006yn}
Abazajian, K. \& Koushiappas, S.~M. 2006, Phys. Rev., D74, 023527,
  astro-ph/0605271

\bibitem[{Abazajian \& Fuller(2002)}]{Abazajian:2002yz}
Abazajian, K.~N. \& Fuller, G.~M. 2002, Phys. Rev., D66, 023526,
  astro-ph/0204293

\bibitem[{Aguilar-Arevalo {et~al.}(2007)}]{AguilarArevalo:2007it}
Aguilar-Arevalo, A.~A. {et~al.} 2007, Phys. Rev. Lett., 98, 231801,
  arXiv:0704.1500

\bibitem[{Asaka {et~al.}(2006)Asaka, Laine, \& Shaposhnikov}]{Asaka:2006rw}
Asaka, T., Laine, M., \& Shaposhnikov, M. 2006, JHEP, 06, 053, hep-ph/0605209

\bibitem[{Biermann \& Kusenko(2006)}]{Biermann:2006bu}
Biermann, P.~L. \& Kusenko, A. 2006, Phys. Rev. Lett., 96, 091301,
  astro-ph/0601004

\bibitem[{Bode {et~al.}(2001)Bode, Ostriker, \& Turok}]{Bode:2000gq}
Bode, P., Ostriker, J.~P., \& Turok, N. 2001, Astrophys. J., 556, 93,
  astro-ph/0010389

\bibitem[{Boyarsky {et~al.}(2006{\natexlab{a}})Boyarsky, Neronov, Ruchayskiy,
  \& Shaposhnikov}]{Boyarsky:2005us}
Boyarsky, A., Neronov, A., Ruchayskiy, O., \& Shaposhnikov, M.
  2006{\natexlab{a}}, Mon. Not. Roy. Astron. Soc., 370, 213, astro-ph/0512509

\bibitem[{Boyarsky {et~al.}(2006{\natexlab{b}})Boyarsky, Neronov, Ruchayskiy,
  \& Shaposhnikov}]{Boyarsky:2006zi}
---. 2006{\natexlab{b}}, Phys. Rev., D74, 103506, astro-ph/0603368

\bibitem[{Boyarsky {et~al.}(2006{\natexlab{c}})Boyarsky, Neronov, Ruchayskiy,
  Shaposhnikov, \& Tkachev}]{Boyarsky:2006fg}
Boyarsky, A., Neronov, A., Ruchayskiy, O., Shaposhnikov, M., \& Tkachev, I.
  2006{\natexlab{c}}, Phys. Rev. Lett., 97, 261302, astro-ph/0603660

\bibitem[{Boyarsky {et~al.}(2007)Boyarsky, Nevalainen, \&
  Ruchayskiy}]{Boyarsky:2006ag}
Boyarsky, A., Nevalainen, J., \& Ruchayskiy, O. 2007, Astron. Astrophys., 471,
  51, astro-ph/0610961

\bibitem[{Bullock {et~al.}(2000)Bullock, Kravtsov, \&
  Weinberg}]{Bullock:2000wn}
Bullock, J.~S., Kravtsov, A.~V., \& Weinberg, D.~H. 2000, Astrophys. J., 539,
  517, astro-ph/0002214

\bibitem[{Dalcanton \& Hogan(2001)}]{Dalcanton:2000hn}
Dalcanton, J.~J. \& Hogan, C.~J. 2001, Astrophys. J., 561, 35, astro-ph/0004381

\bibitem[{Dodelson \& Widrow(1994)}]{Dodelson:1993je}
Dodelson, S. \& Widrow, L.~M. 1994, Phys. Rev. Lett., 72, 17, hep-ph/9303287

\bibitem[{Dolgov \& Sommer-Larsen(2001)}]{Dolgov:2001nq}
Dolgov, A.~D. \& Sommer-Larsen, J. 2001, Astrophys. J., 551, 608

\bibitem[{Goerdt {et~al.}(2006)Goerdt, Moore, Read, Stadel, \&
  Zemp}]{Goerdt:2006rw}
Goerdt, T., Moore, B., Read, J.~I., Stadel, J., \& Zemp, M. 2006, Mon. Not.
  Roy. Astron. Soc., 368, 1073, astro-ph/0601404

\bibitem[{Governato {et~al.}(2004)}]{Governato:2002cv}
Governato, F. {et~al.} 2004, Astrophys. J., 607, 688, astro-ph/0207044

\bibitem[{Hidaka \& Fuller(2007)}]{Hidaka:2007se}
Hidaka, J. \& Fuller, G.~M. 2007, Phys. Rev., D76, 083516, arXiv:0706.3886

\bibitem[{Jena {et~al.}(2005)}]{Jena:2004fc}
Jena, T. {et~al.} 2005, Mon. Not. Roy. Astron. Soc., 361, 70, astro-ph/0412557

\bibitem[{Klypin {et~al.}(1999)Klypin, Kravtsov, Valenzuela, \&
  Prada}]{Klypin:1999uc}
Klypin, A.~A., Kravtsov, A.~V., Valenzuela, O., \& Prada, F. 1999, Astrophys.
  J., 522, 82, astro-ph/9901240

\bibitem[{Komatsu {et~al.}(2008)}]{Komatsu:2008hk}
Komatsu, E. {et~al.} 2008, arXiv:0803.0547

\bibitem[{Kusenko(2006)}]{Kusenko:2006rh}
Kusenko, A. 2006, Phys. Rev. Lett., 97, 241301, hep-ph/0609081

\bibitem[{Kusenko {et~al.}(2008)Kusenko, Mandal, \& Mukherjee}]{Kusenko:2008gh}
Kusenko, A., Mandal, B.~P., \& Mukherjee, A. 2008, Phys. Rev., D77, 123009,
  arXiv:0801.4734

\bibitem[{Kusenko \& Segre(1999)}]{Kusenko:1998bk}
Kusenko, A. \& Segre, G. 1999, Phys. Rev., D59, 061302, astro-ph/9811144

\bibitem[{Loewenstein {et~al.}(2008)Loewenstein, Kusenko, \&
  Biermann}]{Loewenstein:2008yi}
Loewenstein, M., Kusenko, A., \& Biermann, P.~L. 2008, arXiv:0812.2710

\bibitem[{Moore {et~al.}(1999)}]{Moore:1999wf}
Moore, B. {et~al.} 1999, Astrophys. J., 524, L19, astro-ph/9907411

\bibitem[{Narayanan {et~al.}(2000)Narayanan, Spergel, Dave, \&
  Ma}]{Narayanan:2000tp}
Narayanan, V.~K., Spergel, D.~N., Dave, R., \& Ma, C.-P. 2000, Astrophys. J.,
  543, L103, astro-ph/0005095

\bibitem[{O'Shea \& Norman(2006)}]{O'Shea:2006tp}
O'Shea, B.~W. \& Norman, M.~L. 2006, Astrophys. J., 648, 31, astro-ph/0602319

\bibitem[{Pal \& Wolfenstein(1982)}]{Pal:1981rm}
Pal, P.~B. \& Wolfenstein, L. 1982, Phys. Rev., D25, 766

\bibitem[{Peebles(2001)}]{Peebles:2001nv}
Peebles, P. J.~E. 2001, astro-ph/0101127

\bibitem[{Riemer-Sorensen {et~al.}(2006)Riemer-Sorensen, Hansen, \&
  Pedersen}]{Riemer-Sorensen:2006fh}
Riemer-Sorensen, S., Hansen, S.~H., \& Pedersen, K. 2006, Astrophys. J., 644,
  L33, astro-ph/0603661

\bibitem[{Seljak {et~al.}(2006)Seljak, Makarov, McDonald, \&
  Trac}]{Seljak:2006qw}
Seljak, U., Makarov, A., McDonald, P., \& Trac, H. 2006, Phys. Rev. Lett., 97,
  191303, astro-ph/0602430

\bibitem[{Shaposhnikov \& Tkachev(2006)}]{Shaposhnikov:2006xi}
Shaposhnikov, M. \& Tkachev, I. 2006, Phys. Lett., B639, 414, hep-ph/0604236

\bibitem[{Shi \& Fuller(1999)}]{Shi:1998km}
Shi, X.-d. \& Fuller, G.~M. 1999, Phys. Rev. Lett., 82, 2832, astro-ph/9810076

\bibitem[{Strigari {et~al.}(2006)}]{Strigari:2006ue}
Strigari, L.~E. {et~al.} 2006, Astrophys. J., 652, 306, astro-ph/0603775

\bibitem[{Tegmark {et~al.}(2004)}]{Tegmark:2003ud}
Tegmark, M. {et~al.} 2004, Phys. Rev., D69, 103501, astro-ph/0310723

\bibitem[{Tremaine \& Gunn(1979)}]{Tremaine:1979we}
Tremaine, S. \& Gunn, J.~E. 1979, Phys. Rev. Lett., 42, 407

\bibitem[{Watson {et~al.}(2006)Watson, Beacom, Yuksel, \&
  Walker}]{Watson:2006qb}
Watson, C.~R., Beacom, J.~F., Yuksel, H., \& Walker, T.~P. 2006, Phys. Rev.,
  D74, 033009, astro-ph/0605424

\bibitem[{Wilkinson {et~al.}(2006)}]{Wilkinson:2006qq}
Wilkinson, M.~I. {et~al.} 2006, astro-ph/0602186

\bibitem[{Zentner \& Bullock(2002)}]{Zentner:2002xt}
Zentner, A.~R. \& Bullock, J.~S. 2002, Phys. Rev., D66, 043003,
  astro-ph/0205216

\bibitem[{{Zwicky}(1933)}]{zwicky1933}
{Zwicky}, F. 1933, Helvetica Physica Acta, 6, 110

\end{thebibliography}

\bibliographystyle{apje}

\end{document}